\documentclass[aps,prb,superscriptaddress,twocolumn,showpacs,amsmath,amssymb]{revtex4}

\newcommand{\bq}{{\bf q}}

\usepackage{color}
\usepackage{graphicx}
\usepackage{bm}
\definecolor{Blue}{rgb}{0.00, 0.00, 1.00}
\definecolor{Red}{rgb}{1.00, 0.00, 0.00}

\begin{document}

\title{Magneto-electronic properties of multilayer black phosphorus}

\author{Yongjin Jiang}
\affiliation{Department of Electrical \& Computer Engineering, University of Minnesota, Minneapolis, MN 55455, USA}
\affiliation{Center for Statistical and Theoretical Condensed Matter Physics,
ZheJiang Normal University, Jinhua 321004, People’s Republic of China}
\author{ Rafael Rold\'an}
\affiliation{Instituto de Ciencia de Materiales de Madrid, CSIC, Cantoblanco E28049 Madrid, Spain}
\author{ Francisco Guinea}
\affiliation{School of Physics \& Astronomy, University of Manchester, Oxford Road, Manchester, M13 9PL, UK}
\author{Tony Low}
\email{tlow@umn.edu}
\affiliation{Department of Electrical \& Computer Engineering, University of Minnesota, Minneapolis, MN 55455, USA}

\date{\today}

\begin{abstract}
We examine the electronic properties of 2D electron gas in black phosphorus multilayers in the presence of a perpendicular magnetic field, highlighting the role of in-plane anisotropy on various experimental quantities such as ac magneto-conductivity, screening, and magneto-plasmons. We find that resonant structures in the ac conductivity exhibits a red-shift with increasing doping due to inter-band coupling, $\gamma$.  This arises from an extra correction term in the Landau energy spectrum proportional to $n^2\gamma^2$ ($n$ is Landau index), up to second order in $\gamma$. We found also that Coulomb interaction leads to highly anisotropic magneto-excitons.

\end{abstract}


\maketitle

\section{Introduction}

The successful exfoliation of black phosphorus (BP)\cite{li14,xia1,liu14,koenig14} multilayers has triggered tremendous interests in this material, which is also one of the thermodynamically more stable phases of phosphorus, at ambient temperature and pressure.
In comparison to other 2D materials, such as graphene, hexagonal boron nitride (hBN), and transition metal dichalcogenides, multilayers BP has a direct bandgap which spans from $0.3\,{\rm eV} \sim 1.5\,{\rm eV}$ \cite{rudenko14,castellanos14}, hence making it an excellent candidate for infrared optoelectronics \cite{low14cond,low14photo,buscema14}. Since each BP layer forms a puckered surface due to $sp^3$ hybridization, it also reveals highly anisotropic electrical mobility \cite{morita86,li14,liu14}, linear dichroism in optical absorption spectra\cite{morita86,qiao14,low14cond,xia1,YRK15}, anisotropic excitonic structure\cite{tran14,chaves15} and anisotropic plasmons \cite{low14plas, rodin14}. However, BP is not stable in ambient \cite{favron14, island15},
which might render its electrical properties less than pristine.

Recently, encapsulation of BP with hexagonal boron nitride (hBN)\cite{cao15}, all within a controlled inert atmosphere, has allowed for higher carrier mobility in these BP devices\cite{exp1,exp2,exp3,exp4,cao15,bpqhe}. Similar hBN encapsulation has also been applied to other 2D materials such as graphene\cite{wang13} and transition metal dichalcogenides\cite{cui14} to achieve record mobilities. Indeed, the high quality BP has made possible the first observation of prominent quantum magneto-oscillations in these devices\cite{exp1,exp2,exp3,exp4,cao15} and quantum Hall effect\cite{bpqhe}. Hence, theoretical studies of BP in the presence of a magnetic field has also begun receiving attention\cite{zhou14,pereira15}. 

In this paper, we examine the electronic properties of BP multilayer thin film in perpendicular magnetic field, such as its Landau level spectrum, ac conductivity, screening and its collective electronic excitations. In particular, we emphasize the manifestation of anisotropy in these experimentally observable quantities. We begin with a discussion of the model Hamiltonian used to describe multilayer BP in Section II. This is followed by the study of its electronic subband structure in Section III, and its Landau level spectrum in Section IV. Various experimentally relevant quantities such as ac conductivity, collective excitations and screening in the presence of magnetic field will be discussed in Section V-VII respectively.

\section{Model Hamiltonian}

In multilayer BP, broken translational symmetry in the out-of-plane $z$ direction renders the direct energy gap at the $\Gamma$ point instead of the Z point in bulk. The low energy Hamiltonian description of BP near $\Gamma$ point can be expressed as ${\cal H}={\cal H}_z + {\cal H}_{xy}$, with its out-of-plane and in-plane dynamics taken separately. Here, ${\cal H}_{xy}$ is given by\cite{rodin14h,zhou14},
\begin{eqnarray}
{\cal H}_{xy} = \left(
\begin{array}{cc}
E_c + \eta_c k_x^2 + \nu_c k_y^2 & \gamma k_x \\
\gamma k_x  & E_v - \eta_v k_x^2 - \nu_v k_y^2
\end{array} \right)
\label{hamil}
\end{eqnarray}
where $\eta_{c,v}$ and $\nu_{c,v}$ are the respective band parameters, while $\gamma$ describe the effective coupling between the conduction and valence bands.  $E_c$ and $E_v$ denotes the energies of the bulk conduction and valence band edges. We discuss our choice of these band parameters below.


Cyclotron resonance experiments on bulk BP \cite{narita83} found an out-of-plane electron and hole effective masses considerably smaller than that of layered tansition metal dichalcogenides materials  \cite{mattheiss73}. In this work, we adopt an average of experimental \cite{narita83} and theoretically \cite{narita83,low14cond} predicted out-of-plane masses i.e. $m_{cz}\approx 0.2\,m_0$ and $m_{vz}\approx 0.4\,m_0$, $m_0$ being the electron mass. The out-of-plane Hamiltonian is given by,
\begin{eqnarray}
{\cal H}_{z} = -\frac{\hbar^2}{2}\left(
\begin{array}{cc}
m_{cz}^{-1}\partial_z^2 & 0 \\
0  & -m_{vz}^{-1}\partial_z^2
\end{array} \right)+eV(z)
\label{hamilz}
\end{eqnarray}
where $V(z)$ describes the out-of-plane electrostatic potential, typically induced by a bottom metal gate\cite{exp1,exp2,exp3,exp4,cao15}. For a finite BP thickness with given $V(z)$, Eq.\,(\ref{hamilz}) can be diagonalized numerically, leading to electronic subband structure. We denote these subband eigen-energies as $\delta E_{c,v}^j$, and their eigen-functions as $\phi_{c,v}^j(z)$, where $j$ is the subband index.

Close to the $\Gamma$ point, the in-plane band parameters, i.e. $\eta_{c,v}$, $\nu_{c,v}$ and $\gamma$, are related to the in-plane effective masses via\cite{rodin14h},
\begin{eqnarray}
m_{(c,v)x}=\frac{\hbar^2}{\pm 2(\gamma^2/E_g + \eta_{(c,v)})} &\mbox{   ,   }& m_{(c,v)y}=\frac{\hbar^2}{2\nu_{(c,v)}}
\end{eqnarray}
The band parameters $\eta_{c,v}$, $\nu_{c,v}$ and $\gamma$ are chosen such that they yield the known effective masses in the bulk BP limit i.e. $m_{cx}=m_{vx}=0.08\,m_0$, $m_{cy}=0.7\,m_0$ and $m_{vy}=1.0\,m_0$\cite{morita86,narita83}, and $m_{cx}=m_{vx}\approx 0.15\,m_0$ for monolayer BP\cite{rodin14}. $E_g$ is the electronic bandgap of the BP multilayer. The energy gap for bulk BP is $0.3\,$eV\cite{morita86}. While monolayer BP has not been ascertain experimentally, \emph{ab initio} calculation based on the GW method suggests an energy gap of $\sim 1.5-2\,$eV \cite{tran14,rudenko14}. In sum, the band parameters used in this work are; $\eta_c=\eta_v\approx 19.1\,$eV\AA$^2$, $\nu_c\approx 5.45\,$eV\AA$^2$, $\nu_v \approx 3.81\,$eV\AA$^2 $ and  $\gamma=2.84\,$eV\AA. 

 \begin{figure}[t]
   \centering
	\scalebox{0.6}[0.6]{\includegraphics*[viewport=190 270 580 480]
	{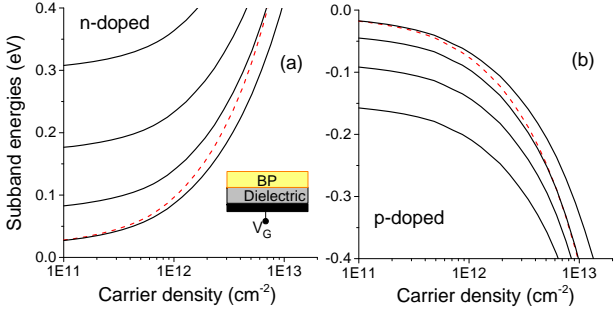}}
 \caption{Electronic subband structure of $10\,$nm BP film 
 for electron (a) and hole (b), showing the lowest four subbands, $\delta E_{c,v}^j$. The red dashed lines corresponds to the Fermi energy $E_f$. These calculations assume $T=300\,$K.}
 \label{fig1}
\end{figure}

\section{Subband Structure}\label{Sec:Subband}

Typical experimental device structure\cite{exp1,exp2,exp3,exp4,cao15} consists of multilayer BP on an insulating dielectric film on a back substrate, which serves also as a back gate, as sketched in the inset of Fig.\,\ref{fig1}(a). The electron density in BP along $z$, $n(z)$, can be obtained from\cite{stern67}, 
\begin{eqnarray}
\nonumber
n(z)
=\frac{k_B T}{\pi\hbar^{2}}\sum_{j}g_{s}m_{cd}\times\\
\mbox{ln}\left[\mbox{exp}\left(\frac{E_{f}-E_{c}-\delta E_{c}^j}{kT}\right)+1\right]\left|\phi_{c}^j(z)\right|^{2}
\label{nden}
\end{eqnarray}
where $g_s$ is the spin degeneracy, $k_B$ is the Boltzmann constant, $E_f$ is the Fermi level, $T$ is the temperature, and $m_{cd}$ refers to the density-of-states mass given by $\sqrt{m_{cx}m_{cy}}$.  Solving Eq.\,(\ref{hamilz}) and the Poisson equation self-consistently, one can then arrives at the numerical solution for the BP electrostatics, an approach well-known in the context of semiconductor inversion layer\cite{stern72}.

The electron and hole subbands, $\delta E_{c,v}^j$, and the Fermi energy $E_f$, are plotted as function of carrier density $n$ in Fig.\,\ref{fig1}(a)-(b) respectively. These calculation assumes $T=300\,$K. The results indicate that across a wide range of carrier densities, only the first subband is occupied, and onset of second subband occupation takes place only when $n\geq 7\times 10^{12}\,$cm$^{-2}$ and $n\geq 5\times 10^{12}\,$cm$^{-2}$ for electron and hole respectively. This  is consistent with recent experimental observations\cite{exp1,exp2,exp3,exp4,cao15} of a 2D electron gas in the quantum limit. Certainly, the transition to multi-subband occupation depends also on the BP thickness.

\begin{figure}[t]
\centering
	\scalebox{0.48}[0.48]{\includegraphics*[viewport=110 90 610 490]
	{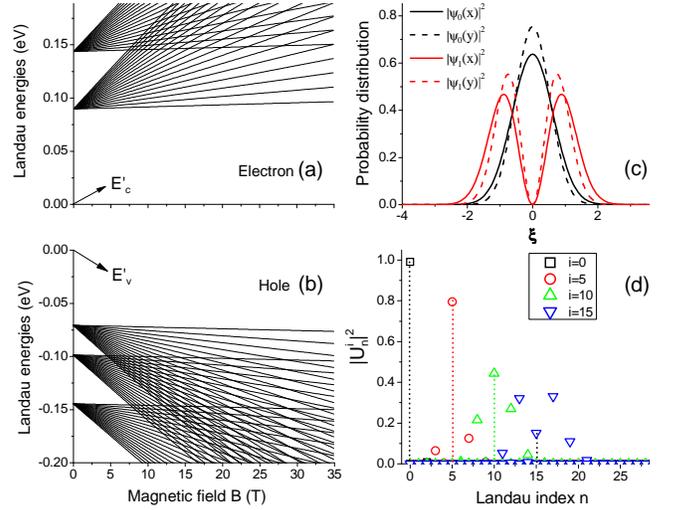}}
\caption{Landau level spectrum for the conduction (a) and valence (b) subbands, calculated for a $10\,$nm thick BP film. The underlying subband structure corresponds to the case with a carrier density of $1\times 10^{12}\,$cm$^{-2}$. In (c), we show the probability distribution $\psi(\xi=\sqrt{\alpha}x,\sqrt{\alpha}y)$ for the two lowest Landau levels in conduction band under different Landau gauges at $B=10\,$T, where $\alpha=\omega_c/2\hbar\eta_c$. In (d), the probability distribution of the $i=0,\,5,\,10,\,15$ eigenstates over the bare oscillator eigenstates $\vert n \rangle$ of lowest conduction subband is plotted.}
\label{fig2}
\end{figure}

\section{Landau level Spectrum}

When an uniform magnetic field $B\hat{z}$ is applied perpendicular to the plane, we have $\vec{p}=\hbar\vec{k}\rightarrow \vec{\Pi}=\vec{p}+e\vec{A}$ by Peierls substitution. With the choice of Landau gauge $\vec{A}=Bx\hat{y}$, we can write
\begin{eqnarray}
\Pi_x=p_x, \Pi_y=p_y+exB,
\end{eqnarray}
which obey the commutation relation  $[\Pi_x,\Pi_y]=-i\hbar eB$. It is useful to define,
\begin{eqnarray}
a=\frac{1}{\sqrt{\hbar\omega_c}}[\sqrt{\eta_c}\Pi_x-i\sqrt{\nu_c}\Pi_y]
\end{eqnarray}
where $\omega_c=2eB\sqrt{\eta_c\nu_c}$. It is easy to verify that $[a,a^\dagger]=1$, and
\begin{eqnarray}
a^{\dagger}a=\frac{1}{\hbar\omega_c}[\eta_c \Pi_x^2+ \nu_c \Pi_y^2]-\frac{1}{2}.
\end{eqnarray}

Introducing the parameters $s=\eta_v/\eta_c$,  $t=\nu_v/\nu_c$ and  $r=s/t$, we can express the dispersing term in valence band as:
\begin{eqnarray}
\eta_v\Pi_x^2+\nu_v\Pi_y^2=s(\eta_c\Pi_x^2+\nu_c\Pi_y^2)+(1-r)\nu_v\Pi_y^2
\end{eqnarray}
which allows us to obtain the effective Hamiltonian under magnetic field $B$ for each pair of electron-hole $j$ subbands,
 \begin{eqnarray}
 \nonumber
{\cal H}^j=\hbar\omega_c\left(%
\begin{array}{cc}
    \epsilon_c^j +a^{\dagger}a+\frac{1}{2}&\tilde{\gamma}(a^{\dagger}+a)\\
    \tilde{\gamma}(a^{\dagger}+a) & \epsilon_v^j  +s(a^{\dagger}a+\frac{1}{2})+\tilde{r}(a^{\dagger}-a)^2\\
\end{array}%
\right)\\
\label{Heff}
\end{eqnarray}
where $\tilde{\gamma}=\frac{\gamma}{2\sqrt{\hbar \omega_c\eta_c}} $, $\tilde{r}=-\frac{1}{4}t(1-r)$ and $\epsilon_{c,v}^j=(E'_{c,v} + \delta E_{c,v}^j)/\hbar\omega_c$. We note that, for the representative numerical results presented here, the band edges $E_{c,v}$ are adjusted so as to reproduce the estimated electronic bandgap of $\approx 0.5\,$eV\cite{tran14} for a $10\,$nm BP film, i.e. $(E'_{c}+\delta E_c^1)-(E'_{v}+\delta E_v^1)=0.5\,$eV.

Now the matrix Hamiltonian is dimensionless, and the eigenvalue problem can be solved numerically. The procedure is as follows. First, we assume the following ansatz for the eigenvector of Eq.\,(\ref{Heff}),
\begin{eqnarray}
\psi(x)=\left(%
\begin{array}{c}
    U(x)\\
    V(x)\\
\end{array}%
\right)=\left(%
\begin{array}{c}
    \sum_{n=0}^{n_{max}} U_n\phi_n(\alpha (x-x_0))\\
    \sum_{n=0}^{n_{max}} V_n\phi_n(\alpha (x-x_0))\\
\end{array}%
\right).
\label{eigenfunction}
\end{eqnarray}
where $n_{max}$ sets the truncation of the expansion and $\phi_n(\alpha (x-x_0))$ is the $n$'th eigenstate of harmonic oscillator centered at $x_0=\dfrac{p_y}{eB}$ with $\alpha=\dfrac{\omega_c}{2\hbar\eta_c}$. Note the wave function will include $e^{ip_yy/\hbar}$ factor, associated to the good quantum number $p_y$.  We can determine the coefficients  $U_n$ and $V_n$ from the energy eigen-problem ${\cal H}^j\psi=E\psi$. Explicitly, we have the following  relations
\begin{widetext}
\begin{eqnarray}
\begin{array}{c}
    (\Delta_1+n-\tilde{E})U_n+\tilde{\gamma}(\sqrt{n}V_{n-1}+\sqrt{n+1}V_{n+1})=0\\
    \tilde{\gamma}(\sqrt{n}U_{n-1}+\sqrt{n+1}U_{n+1})+     (\Delta_2+\tilde{s}n-\tilde{E})V_n+\tilde{r}(\sqrt{n(n-1)}V_{n-2}+\sqrt{(n+2)(n+1)}V_{n+2})=0 \\
\end{array}%
\label{eigen}
\end{eqnarray}
\end{widetext}
where $\Delta_1=\epsilon_{c}^j+\frac{1}{2} $, $\Delta_2=\epsilon_{v}^j+\frac{s}{2}-\tilde{r}$, $\tilde{s}=s-2\tilde{r}$ and  
 $\tilde{E}=E/\hbar \omega_c$, $E$ being the energy of  Landau level corresponding to the eigen-function (\ref{eigenfunction}).    
Numerically, we need to introduce a truncation condition, which  we  set to $U_{n>n_{max}}=V_{n>n_{max}}=0$. 

\textcolor{black}{It is not difficult to check that for most physical relevant cases, $\Delta_{1,2}\gg n,\tilde{\gamma}$, which justifies a perturbative consideration for Eq.\,(\ref{eigen}). For the electron (hole) spectrum, we can write,
$\tilde{E}^{e/h}=\tilde{E}_{0}^{e/h}+\delta \tilde{E}^{e/h}$ with $\tilde{E}_{0}^{e}=\Delta_1+n$, $\tilde{E}_{0}^{h}=\Delta_2+\tilde{s}n$,  and the second order perturbations terms are:
\begin{eqnarray}
\delta \tilde{E}^{e/h}= \pm\tilde{\gamma}^2(c_0+c_1n+c_2n^2)
\label{E2perturbation}
\end{eqnarray}
where $c_0=\dfrac{1}{(\Delta_1-\Delta_2)}$, $c_1=\dfrac{2}{(\Delta_1-\Delta_2)}-\dfrac{1-\tilde{s}}{(\Delta_1-\Delta_2)^2}$ and $c_2=\dfrac{-2(1-\tilde{s})}{(\Delta_1-\Delta_2)^2}$.
From Eq.(\ref{E2perturbation}), we may understand the deviation of resonant frequency of the ac conductivity from conventional 2D electron gas case, to be discussed in Section V.}

\textcolor{black}{From numerical recipe described below Eq.(\ref{eigen})}, we obtain the Landau level spectrum, as shown in Fig.\,\ref{fig2}(a)-(b) for conduction and valence bands respectively. The dispersion is typical of 2D electron gas system, exhibiting linear dependence with $B$. However, there are differences due to the finite inter-band coupling $\gamma$ and anisotropy of BP, which will be discussed later.  In the calculations, we set the doping to be $1\times 10^{12}\,$cm$^{-2}$. We note that crossing of each Landau level acquires additional carrier density of $\Delta n = eBg/h\sim \tfrac{1}{2}B\times 10^{11}\,$cm$^{-2}$, where $g=2$ is the spin degeneracy. Hence, at the assumed doping of $1\times 10^{12}\,$cm$^{-2}$ and $B=10\,$T, the filling factor is $2$. For carrier densities larger than $1\times 10^{13}\,$cm$^{-2}$, the filling factor can be as large as $20$. It might then be possible to observe multi-subband phenomena, especially for the hole case. For the experimental doping range, only the Landau levels of the lowest two subbands are physically relevant. 


The anisotropy of the problem is encoded in the wavefunctions. In Fig.\,\ref{fig2}(c) we plot the wavefunction probability along the two in-plane spatial coordinates for the first two Landau levels, expressed in their dimensionless coordinate (i.e., $\xi=\sqrt{\alpha}x$, $\xi=\sqrt{\alpha}y$). Two respective gauges are used, i.e., for gauges $\vec{A}=Bx\hat{y}$ and $\vec{A}=-By\hat{x}$, we get $\psi(x)$ and $\psi(y)$ respectively.
Due to the anisotropy inherent in the model Hamiltonian, we can clearly discern the difference between probability distribution over these two gauges. We will show that this anisotropy in wave function can result in prominent anisotropy in various experimental quantities such as ac conductivities and magneto-plasmons. In  Fig.\,\ref{fig2}(d), the probability distribution over the eigenstates $\vert n\rangle$ of the \textit{bare} harmonic oscillator which forms the upper diagonal terms in Eq.\,(\ref{Heff}) is plotted. With the increase of Landau level, the probability distribution over $\vert n\rangle$ becomes more broadened. This can be understood from perturbation point of view, i.e.,  the matrix elements quantifying the perturbation upon the \textit{bare} eigenstates increases with factors $\sqrt{n}$'s. 

\textcolor{black}{Before concluding this section,  it is interesting to compare Landau levels in BP with the other electron gas system e.g. conventional 2D electron gas (i.e. Schr\"odinger fermions) and graphene (i.e. Dirac fermions), summarized as follows:
\begin{eqnarray}
E_n=\begin{cases} \hbar\omega_c(n+\tfrac{1}{2}):    \mbox{Schr\"odinger fermions}
 \\ \mbox{sgn}(n)v_f\sqrt{2e\hbar B|n|}:   \mbox{Dirac fermions}
 \\ \hbar\omega_c(n+\tfrac{1}{2}+\tilde{\gamma}^2(c_0+c_1n+c_2n^2)):  \mbox{BP} \end{cases}
 \label{Landaulevelcompare}
\end{eqnarray}}
\textcolor{black}{The BP effective Hamiltonian, described by Eq.(\ref{hamil}), have features not embodied in Schr\"odinger fermion description in the finer energy scale proportional to $\gamma^2$, and have experimental consequences discussed in Section V.   
We note also that compared to other gapped Dirac systems, such as gaped graphene or transition metal dichalcogenides, BP's gap is placed at the time-reversal invariant $\Gamma$ point instead of the inequivalent $K$ and $K'$ points of the BZ. Hence, its Landau level spectrum resembles more to that of Schr\"odinger fermions rather than that of massive Dirac fermions.\cite{Goerbig14}}



\section{ac magneto-conductivity}
 In this section, we study the ac magneto-conductivity $\sigma_{\alpha \beta}$ of the $10\,$nm BP multilayer film. Having the energy spectrum and wave-functions of Landau levels at hand, as presented in the previous section, we can numerically calculate the $\sigma_{\alpha \beta}$'s directly. According to the Kubo formula\cite{Hbert}, we have:

 \begin{eqnarray}
\sigma_{\alpha \beta}(\omega)&=&\frac{-ie^2\hbar }{2\pi l_B^2}\sum_{snjs'n'j'}\frac{f(E_{snj})-f(E_{s'n'j'})}{E_{snj}-E_{s'n'j'}}\nonumber\\
&\times& \frac{\left \langle \Phi_{snj} \left|\hat{v}_{\alpha} \right| \Phi_{s'n'j'} \right \rangle \left \langle \Phi_{s'n'j'} \left|\hat{v}_{\beta} \right| \Phi_{snj} \right \rangle }{\hbar\omega-  E_{snj}+E_{s'n'j'}+i\Gamma}
\label{eq4}
\end{eqnarray}
where the velocities are defined by $v_i=\frac{\partial H}{\partial p_i}$. Explicitly, we have:
   \begin{eqnarray}
   \begin{array}{c}  
v_x=\left(%
\begin{array}{cc}
    \sqrt{\hbar\omega_c\eta_c}(a+a^{\dagger})&\gamma\\
    \gamma& \eta_{v}\sqrt{\hbar\omega_c/\eta_c}(a+a^{\dagger})\\
\end{array}%
\right)\\
v_y=\left(%
\begin{array}{cc}
    \sqrt{\hbar\omega_c\nu_c}(a^{\dagger}-a)/i& 0\\
    0&  \nu_v\sqrt{\hbar\omega_c/\nu_c}(a^{\dagger}-a)/i\\
\end{array}%
\right)\\
\end{array}
\label{velocity}
\end{eqnarray}

\begin{figure}[t]
\centering
	\scalebox{0.6}[0.6]{\includegraphics*[viewport=155 110 600 500]
	{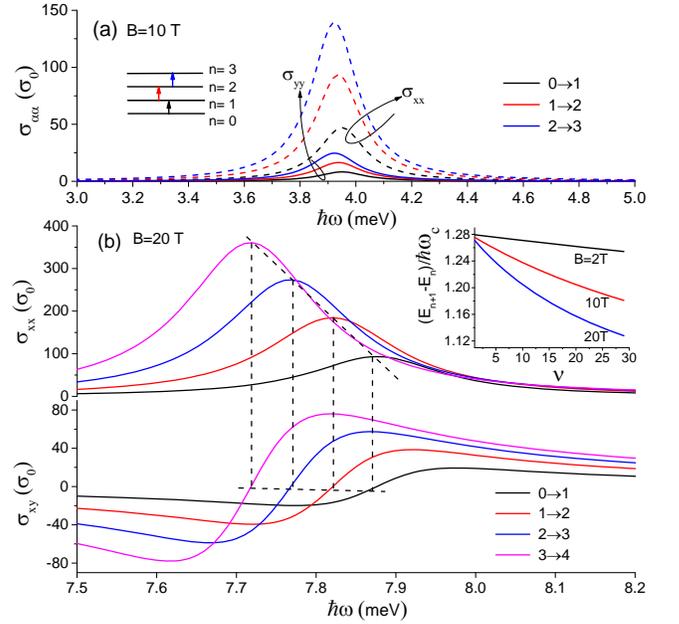}}
\caption{ac conductivities as function of frequency $\omega$. The anisotropy between $\sigma_{xx}$ and $\sigma_{yy}$, at $B=10\,$T is displayed in (a), for different $n\rightarrow n+1$ transitions. Zooming in, we can observe slight shift of the resonant frequency for $\sigma_{xx}$  and $\sigma_{xy}$, as shown in (b), calculated for $B=20\,$T. The damping constant is set to be $\Gamma=10^{-4}\,$eV and temperature $T=10\,$K. The inset of (b) shows how the resonant frequency shift depends on filling factor $\nu$ and $B$. Conductivity are in units of $\sigma_0=e^2/\hbar$.}
\label{fig3}
\end{figure}

Our results for ac conductivity as a function of frequency are presented in Fig.\,\ref{fig3}. The ac longitudinal magneto-conductivities for filling factor $\nu=$ 1 to 3 (electron doped) is shown in Fig.\,\ref{fig3}(a). The inset depicts the transitions between the nearest Landau levels
for cases with $\nu=1$ to $3$, i.e. the resonance for each case corresponds to a particular transition process $\vert n\rangle$ to $\vert n+1\rangle$, and occurs at the terahertz frequencies.
 Prominently, $\sigma_{xx}$ is about $5\sim 10$ times larger than $\sigma_{yy}$. 
We note the spatial anisotropy in the wavefunctions of the Landau levels, albeit small, as shown in Fig.\,\ref{fig2}c. The anisotropy in the magneto-optical conductivity tensor arises mainly from the anisotropy in the velocity operators, i.e. $v_x$ and $v_y$, which accounts for the anisotropic optical transitions dipole.  




We should point out here that the resonant structure observed here in BP is more like conventional 2D electron gas rather than in graphene\cite{Giant}, which has multiple resonant structures. Contrary to the conventional case, we find here that the resonant frequency is slightly red-shifted with  increasing doping (or $\nu$). This is shown in Fig.\,\ref{fig3}(b), where the longitudinal conductivity $\sigma_{xx}$ and Hall conductivity $\sigma_{xy}$ are displayed for $\nu=1-4$, calculated for $B=20\,$T. 
This red-shift also increases with magnetic field as depicted in the inset. 
We note that the frequency shift increase in \textit{almost uniform} steps each time the filling factor decreases by $1$, \textcolor{black}{ 
Interestingly, this red-shifting behavior can be understood from the perturbation expression in Eq.\,(\ref{E2perturbation}) rather straightforwardly, from which we arrived at,
\begin{equation}
\tilde{E}^e(n+1)-\tilde{E}^e(n)=1+\tilde{\gamma}^2[c_1+c_2(2n+1)]   
\end{equation}
This expression accounts for linear dependence of the resonance frequency on $n$ quantitatively, and is a direct result of the interband coupling $\gamma$. In other words, this red-shift in ac conductivity can be used to determine $\gamma$ in BP. }
 As depicted in the inset of Fig.\,\ref{fig3}(b), it can be seen that the linear relation holds when $\nu$ is small. We anticipate experimental progress in magnetic oscillations of BP to shed light on this issue\cite{exp1,exp2,exp3,exp4,cao15}. It is also worthy to mention that the longitudinal  and
Hall conductivities calculated here can be directly measured through absorption and Faraday rotation experiments\cite{Giant} via terahertz spectroscopy.

\section{Collective excitations}

\begin{figure}[t]
\centering
 \includegraphics[width=0.45\textwidth]{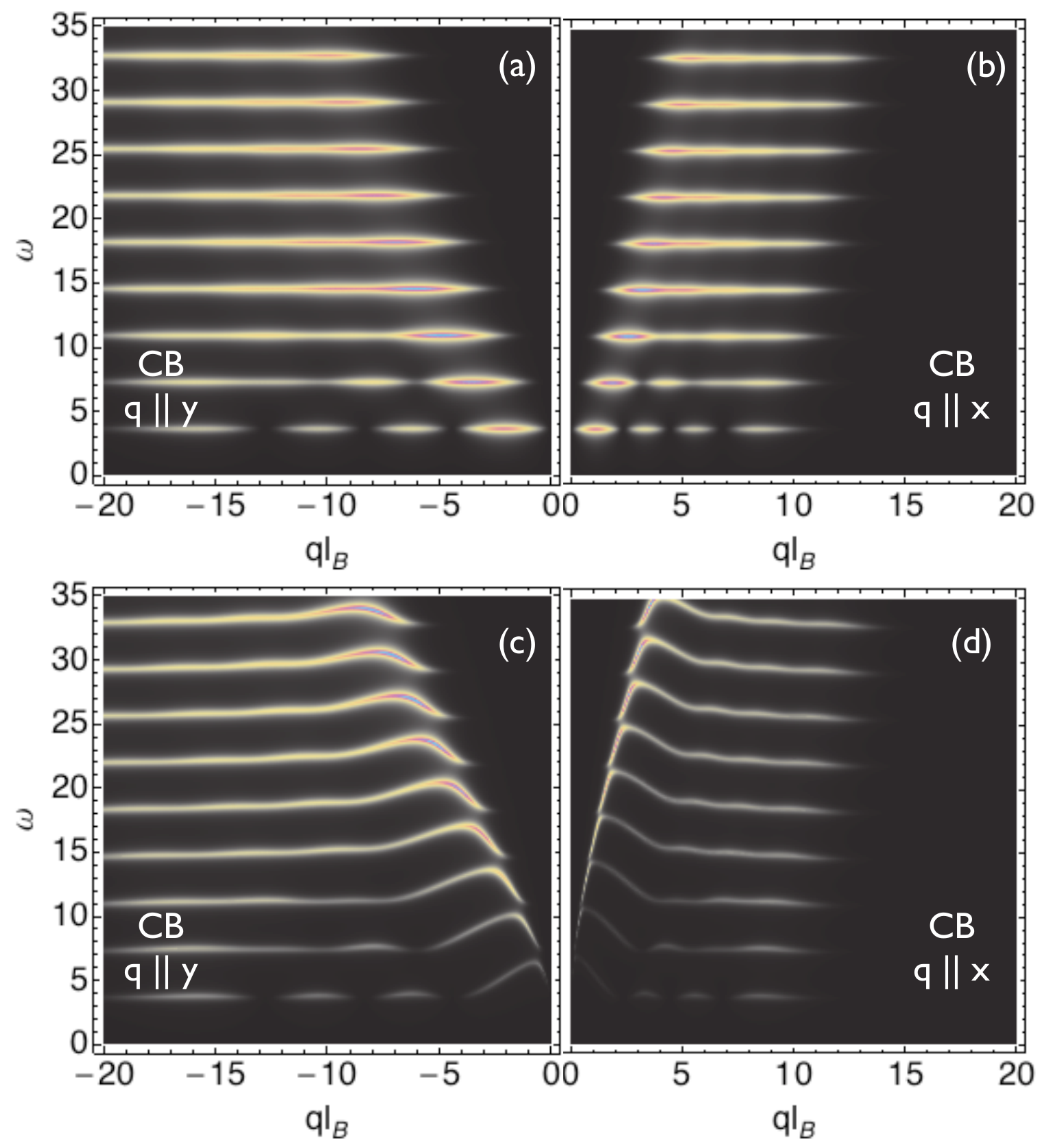}
\caption{Excitation spectrum of BP, as obtained by density plots of ${\rm Im}\Pi(\bq,\omega)$. (a) and (b)
correspond to non-interacting polarization of BP, Eq.
(\ref{Eq:Pi0}), for $q\parallel y$ and $q\parallel x$ respectively. Plots  (c) and (d) include electron-electron
interactions in the RPA. The strength of the electron-electron interaction is chosen to be $r_s\approx 3$, and $N_F=3$.}
  \label{Fig:MP}
\end{figure}

In this section we study the excitation spectrum of BP in the presence of a quantizing magnetic field applied perpendicular to the sample, including the effect of electron-electron interaction. As we have seen in Sec. \ref{Sec:Subband}, a 2DEG is created in a multi-layer BP, with occupation of only the first subband, unless the doping exceeds $\sim 5\times10^{12}~{\rm cm}^{-2}$. Therefore in this section we concentrate on the particle-hole excitation spectrum of the 2DEG formed by the carriers of a 10nm thick BP multi-layer.   In the absence of Coulomb interaction, the particle-hole excitation spectrum, that enclose the region of the $\omega-q$ plane in which it is possible to excite electron-hole pairs, can be calculated from ${\rm Im}\Pi^0(\bq,\omega)\ne 0$, where $\Pi^0(\bq,\omega)$ is the non-interacting polarization function. The bare polarizability of BP in the quantum Hall regime can be expressed in terms of the standard result for a 2DEG\cite{KH84}
\begin{equation}\label{Eq:Pi0}
\Pi^0(\bq,\omega)=\sum_{m=1}^{N_c}{\sum}'\frac{{\cal
F}_{n,m}(\bq)}{\omega-m\omega_c+i\Gamma}+(\omega^+\rightarrow
-\omega^-)
\end{equation}
where $\sum'=\sum_{n=\max(0,N_F-m)}^{N_F}$ and
$\omega^+\rightarrow -\omega^-$ indicates the replacement
$\omega+i\Gamma\rightarrow -\omega-i\Gamma$, and $N_F$ is the index of the last occupied LL. As we have seen before, the anisotropy of the BP band structure is encoded in the wave-function, whose overlaps leads to different form factors in the $x$- and $y$-direction
\begin{equation}\label{FF2DEG}
{\cal F}_{n,m}(\bq)=\frac{e^{-\alpha\frac{q^2l_B^2}{2}}}{2\pi
l_B^2}\frac{n!}{(n+m)!}\left (\alpha\frac{q^2l_B^2}{2}\right)^m
\left[L_n^m\left(\alpha\frac{q^2l_B^2}{2}\right )\right]^2.
\end{equation}
where $l_B=\sqrt{\hbar c/eB}$ is the magnetic length, $\alpha=\omega_c/2\hbar\eta_{(c,v)}$ for the case when $q_x$ is a good quantum number, and $\alpha=\omega_c/2\hbar\nu_{(c,v)}$ when the gauge chosen leads to a good $q_y$ quantum number. We note that the finite interband coupling, $\gamma$, can renormalize\cite{zhou14} the band parameters ($\eta_{(c,v)}$and $\nu_{(c,v)}$) but we checked that the effect is small in this case. A density plot of ${\rm Im}\,\Pi^0$ is shown in Fig. \ref{Fig:MP}(a) and (b) for electron doping with
$N_F=3$. In the presence of a quantizing magnetic field, ${\rm
Im}\,\Pi^0(\bq,\omega)$ is a sum of Lorentzian peaks centered at
$\omega=m\omega_c$, where $m$ is the difference between
the LL indices of the electron $n'$ and the hole $n$, $m\equiv n'-n\ge 1$.\cite{GV05,RGF10} The
excitation spectrum is chopped into horizontal lines, separated by a constant
energy $\omega_c$. The peculiarities of the BP spectrum of Fig. \ref{Fig:MP}(a)-(b), like the presence of a superstructure of $N_F+1$ brighter regions and the nodes of the first horizontal line at $\omega=\omega_c$, are due to the form of the LL wavefunctions and have been studied in detail in Refs. \onlinecite{RFG09,RGF10}. As in the $B=0$ case,\cite{low14plas} anisotropy of BP band structure leads to a wider spectrum in the $q_y$ direction as compared to $q_x$ direction. Whereas we only show results for electron doping in Fig. \ref{Fig:MP}, a similar spectrum is found for the hole doped case, with the difference that the separation between consecutive horizontal lines is narrower for the latter, due to larger effective mass of the valence band as compared to the conduction band.

We next consider the effect of Coulomb interaction within the Random Phase Approximation (RPA), which {\it dresses} the electron-hole polarization function as
\begin{equation}\label{Eq:PiRPA}
\Pi(\bq,\omega)=\frac{\Pi^0(\bq,\omega)}{\varepsilon(\bq,\omega)}
\end{equation}
where the dielectric function is obtained as
\begin{equation}
\varepsilon(\bq,\omega)=1-v(\bq)\Pi^0(\bq,\omega),
\end{equation}
in terms of the non-interacting polarization function $\Pi^0$ and the two-dimensional Coulomb potential
in momentum space
\begin{equation}
v(\bq)=\frac{2\pi e^2}{\kappa |\bq|}
\end{equation}
where $\kappa$ is the background dielectric constant. The strength of the Coulomb interaction is usually expressed in terms of the dimensionless parameter $r_s=2m_be^2/\kappa k_F$, where the band mass in the present case is $m_b=\sqrt{m_xm_y}$, and $k_F=\sqrt{2N_F+1}/l_B=\sqrt{\pi n_{el}}$ in terms of the carrier density $n_{el}$.  Long-range electron-electron interaction leads to the appearance of collective plasmon modes in the spectrum. Their dispersion relation can be calculated from the zeros of the dielectric function. In Fig. \ref{Fig:MP}(c) and (d) we show how the non-interacting electron-hole spectrum [shown in Fig. \ref{Fig:MP}(a) an (b)] is modified due to interactions. The electron-hole horizontal lines of Fig. \ref{Fig:MP}(a) and (b) acquire a dispersion, leading to the so-called magneto-excitons or magneto-plasmons. As in the $B=0$ case,\cite{low14plas} Coulomb interaction leads to highly anisotropic magneto-plasmons, with higher dispersion for $q\parallel x$ than for $q\parallel y$ directions. This is due to the fact that the {\it effective} mass is smaller in the $x$ direction than in the $y$ direction. We also notice that the excitation spectrum of BP greatly differs from that of doped graphene. In fact, the characteristic linear dispersion relation in graphene leads to a {\it relativistic}  quantization of the spectrum into a set of non-equidistant LLs.\cite{G11} As a consequence, long range Coulomb interaction in graphene leads to a set of highly dispersing {\it linear} magneto-plasmons,\cite{RFG09} that differs from the equidistant magnetoexcitons (separated by a well defined cyclotron frequency energy, $\omega_c$) in BP.

\section{Static screening}

\begin{figure}[t]
\centering{
  \includegraphics[width=0.4\textwidth]{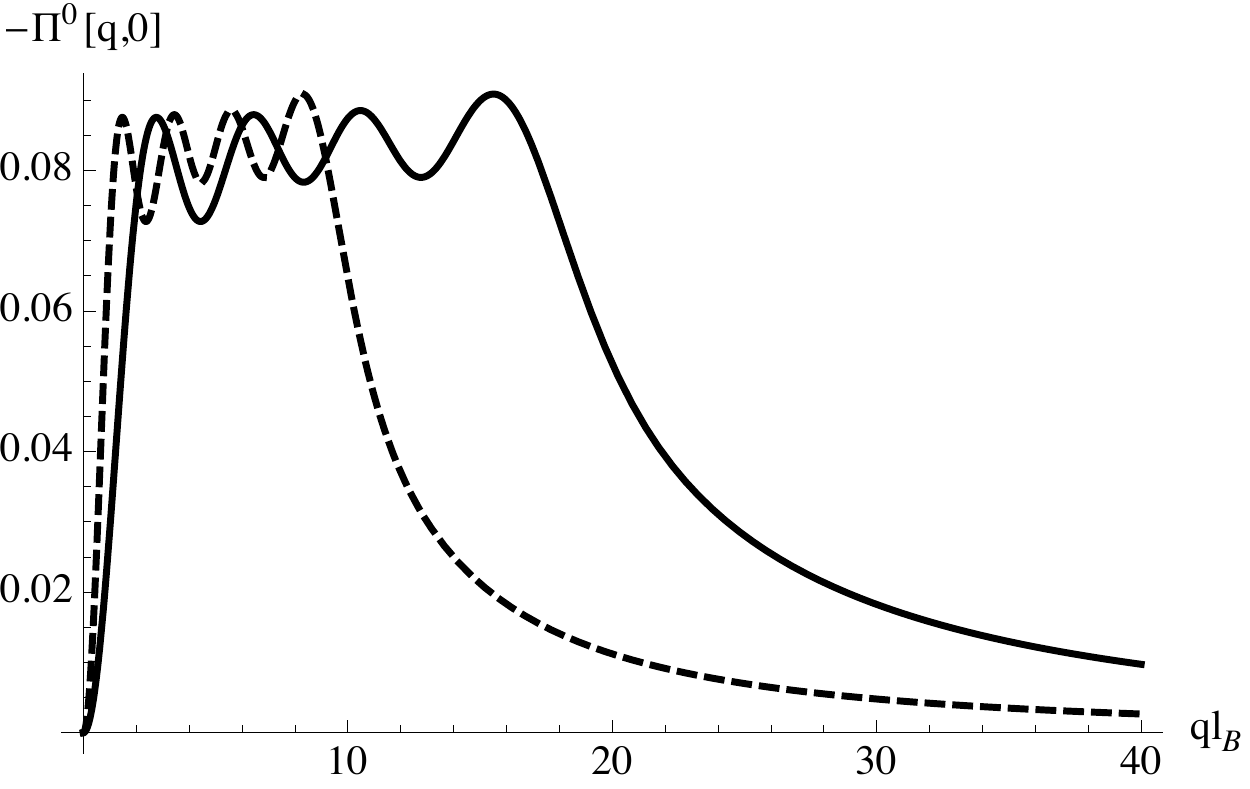}
  }
  \centering{
    \includegraphics[width=0.4\textwidth]{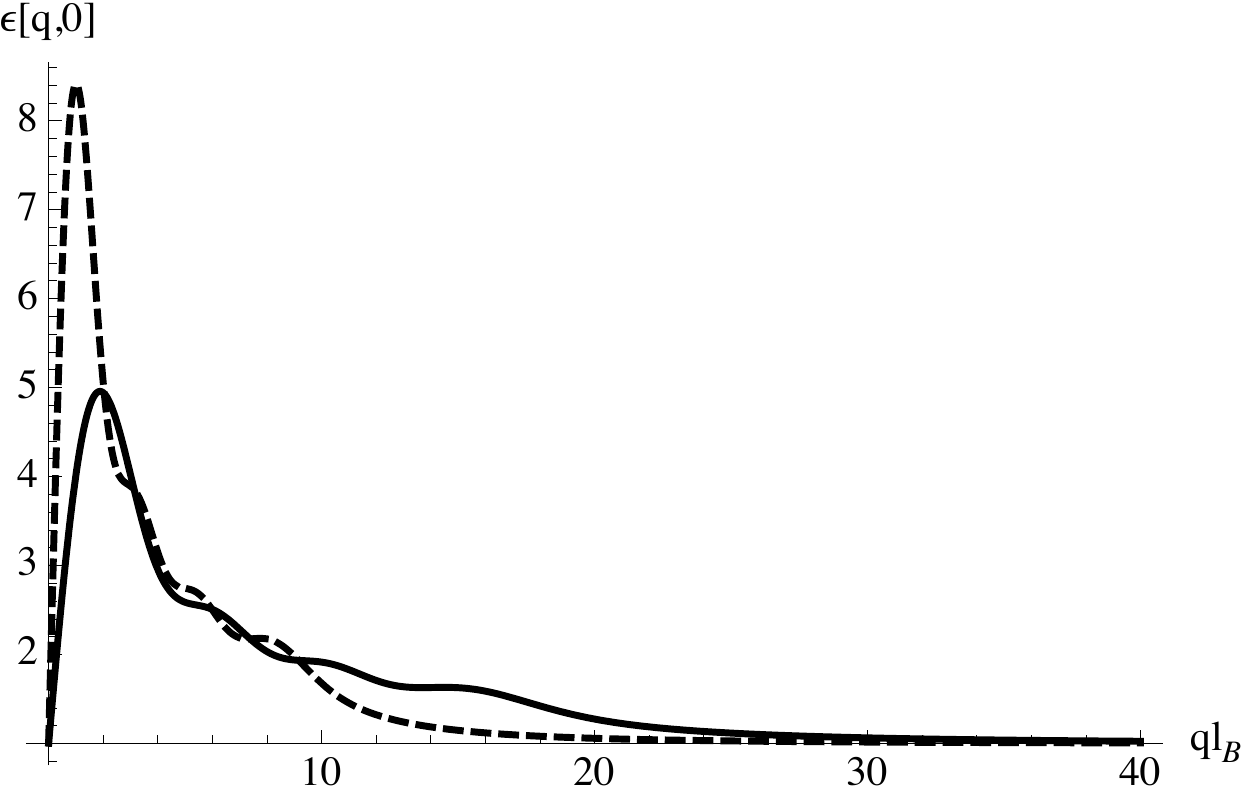}}
\caption{Static screening. Static polarization function $-\Pi^0(\bq,0)$ (top) and dielectric function $\varepsilon(\bq)$ (bottom) for the conduction band with $\bq \parallel y$ (solid lines) and for $\bq \parallel x$ (dashed lines). We have used $N_F=3$. }
  \label{Fig:StaticScreening}
\end{figure}

In this section we focus on the properties of $\Pi^0(\bq)=\Pi^0(\bq,\omega=0)$ and $\varepsilon(\bq)=\varepsilon(\bq,\omega=0)$
in the static limit, for which the polarization function is entirely real.
The polarizability of BP in a strong magnetic field is shown in Fig. \ref{Fig:StaticScreening}(a). One first observe that, as in a standard 2DEG,\cite{GV05} the static polarizability
 tends to zero as $\Pi^0(\bq\rightarrow 0)\propto q^2$ for
$B\neq 0$. The reason for this is that the main contribution to
$\Pi^0(\bq)$ comes from $\bq=0$ excitations in the vicinity of
the Fermi energy $E_f$. This differs from the $B=0$ case, for which the Fermi level cuts the band and  
there are $\bq=0$ excitations. In the integer quantum Hall regime, however,
$E_f$ lies in the cyclotron gap between the highest occupied LL
$N_F$ and the lowest unoccupied one $N_F+1$. Since this energy gap must be
overcome by excitations with $\bq=0$, then its spectral weight tends
to zero. Notice that $\Pi^0(q=0)$ coincides with
the density of states at the Fermi energy because the latter
vanishes for $B> 0$ when $E_f$ lies in the gap. One also notices that the wave-vector at which the polarizability starts  to vanish ($2k_F$) is larger for $\bq \parallel y$ that for $\bq \parallel x$, due to the anisotropy of the BP Fermi surface. One further notices the oscillatory behavior of the static polarizability, below $2k_F$, due to the wave-function overlap between the electron and the hole, leading to $N_F+1$ maxima.\cite{GV05,RGF10}

It is interesting to compare the screening properties of BP and graphene, and we start by briefly discussing the $B=0$ case. For $q\rightarrow 0$ we have $\varepsilon(q)\approx 1+q_{TF}/q$, where $q_{TF}\equiv
2\pi e^2 \rho(E_f)/\kappa$ is the 2D Thomas-Fermi
wave-vector, in terms of the density of states $\rho(E_f)$.  Since the density of states (per unit area) is approximately a constant equal to $g_sm_b/2\pi$ for BP, where $g_s=2$ accounts for the spin degeneracy, whereas for graphene $\rho(E_f)$
is energy dependent and given approximately by $gE_f/(2\pi v_F^2)$, where $v_F$ is the Fermi velocity and
$g=g_sg_v=4$ accounts for spin and valley degeneracy, one obtains a density independent $q_{TF}$ for
BP, whereas it scales as $k_F$ for graphene.  Therefore, in the two cases and for $B=0$, the dielectric function diverges as $\varepsilon \sim
q_{TF}/q\rightarrow \infty$ when $q\to 0$.\cite{low14plas} However, it is important to notice that
the density dependence in the numerator of
$\varepsilon(q\rightarrow 0)$ in graphene implies the
absence of screening in undoped graphene (i.e. for $k_F=0$) at long
distances. For short wavelengths  $q\gg 2k_F$,
$\varepsilon(q)\rightarrow 1$ in a BP, whereas for graphene,
$\varepsilon(q)\rightarrow 1+\pi r_s/2$. The extra
contribution $\pi r_s/2$ to the dielectric function of graphene
at $q\gg 2k_F$ is related to virtual
inter-band particle-hole excitations.\cite{RGF10} In summary, at short
wavelengths and for $B=0$, BP (like a standard 2DEG) does not screen at all ($\varepsilon \to 1$),
whereas  graphene screens as a dielectric due to its filled valence band.

The above picture changes in the presence of a quantizing magnetic field. In
Fig. \ref{Fig:StaticScreening} we have plotted the static polarization and
dielectric functions for BP, respectively. At
long wavelengths, $\varepsilon(q)-1\propto q$. In the limit of $N_F\gg 1$, the known result for a 2DEG and for graphene,\cite{RGF10} $\varepsilon(q)-1\propto r_sN_F^{3/2}ql_B$ as $q\rightarrow 0$, applies also for BP. The difference between graphene and BP is encoded in the density dependence of $r_s$ in the two cases: since $r_s\sim N_F^{-1/2}$ in BP, $\varepsilon$ grows linearly with $N_F$ in this case. However, $r_s\equiv e^2/\kappa v_F$ is density-independent in graphene, leading to a dielectric function proportional to $N_F^{3/2}$.  On the other hand, in both graphene and BP,
$\varepsilon(q,\omega=0)\rightarrow 1$ as $q\rightarrow 0$,
which implies that there is no screening at long distances. Finally, the behavior of the dielectric function in a magnetic field at $q\gg 2k_F$
in BP is similar to the $B=0$  limit, corresponding to the standard metallic like screening governed by intra-band processes.\cite{low14plas}

\section{Conclusion}

In conclusion, we have examined the electronic properties of 2D electron gas in black phosphorus multilayers due to the presence of a perpendicular magnetic field. We highlight in this work the in-plane anisotropy reflected in various experimental quantities such as its ac magneto-conductivity, screening, and collective excitations. We found that resonant structures in the ac conductivity exhibits a red-shift with increasing doping due to interband coupling, suggesting possible electric modulation of light absorption and Faraday rotation. Coulomb interaction also leads to highly anisotropic magneto-plasmons.

\acknowledgments

We thank M. O. Goerbig and J.-N. Fuchs for useful conversations. FG and RR acknowledge support from the Spanish Ministry of Economy (MINECO) through Grant No. FIS2011-23713, the European Research Council Advanced Grant (contract 290846), and the European Commission under the Graphene Flagship, contract CNECT-ICT-604391. R.R. acknowledges financial support from the Juan de la Cierva Program. YJ and TL acknowledge support from University of Minnesota start-up fund. Y.J. acknowledge support from the National Natural Science Foundation of China under Grants No. 11474255 during intial part of the project.

\end{document}